\documentclass[prd,showkeys,floatfix,nofootinbib,showpacs,
               preprint,12pt,
               tightenlines,fleqn]{revtex4-1}

\usepackage{amsmath,amssymb,revsymb,graphicx,dcolumn}
\usepackage{leftidx,slashed,datetime}

\usepackage{mwe}
\newcommand{\imineq}[2]{\vcenter{\hbox{\includegraphics[height=#2ex]{#1}}}}

\newcommand{\beq}{\begin{equation}}
\newcommand{\eeq}{\end{equation}}
\newcommand{\beqa}{\begin{eqnarray}}
\newcommand{\eeqa}{\end{eqnarray}}
\newcommand{\bsubeqs}{\begin{subequations}}
\newcommand{\esubeqs}{\end{subequations}}

\usepackage{marginnote}

\begin{document}
\title[]
      {The Minkowski quantum vacuum does not gravitate} 
\author{Viacheslav A. Emelyanov}
\email{viacheslav.emelyanov@kit.edu}
\affiliation{Institute for Theoretical Physics,\\
Karlsruhe Institute of Technology,\\
76131 Karlsruhe, Germany\\}

\date{\today}

\begin{abstract}
\vspace*{2.5mm}\noindent
We show that a non-zero renormalised value of the zero-point
energy in $\lambda\phi^4$-theory over
Minkowski spacetime is in tension
with the scalar-field equation at two-loop order in perturbation theory.
\end{abstract}


\maketitle

\section{Introduction}

Quantum fields give rise to an infinite vacuum energy density~\cite{Pauli,Zeldovich,Weinberg,Martin},
which arises from quantum-field fluctuations taking place even in the absence
of matter. Yet, assuming that semi-classical quantum field theory is reliable only
up to the Planck-energy scale, the cut-off estimate yields zero-point-energy
density which is finite, though, but in a notorious tension with astrophysical
observations.

In the presence of matter, however, there are quantum effects occurring in
nature, which cannot be understood without quantum-field fluctuations. These
are the spontaneous emission of a photon by excited atoms, the Lamb shift,
the anomalous magnetic moment of the electron, and so forth~\cite{Milonni}.
This means quantum-field fluctuations do manifest themselves in nature and,
hence, the zero-point energy poses a serious problem.

Lorentz symmetry implies that vacuum stress-energy tensor is proportional to
the metric tensor~\cite{Zeldovich}. As a consequence, the vacuum
energy density must equal a quarter of the stress-tensor trace. In the case
of Maxwell theory, the photon field cannot thus give a non-vanishing
Lorentz-invariant vacuum stress tensor, due to conformal invariance of the
theory. Still, its vacuum energy density is, rigorously speaking, infinite. This tension can be
re-solved if the stress-energy tensor is (properly) regularised and then
renormalised to zero. This procedure does not result in a fine-tuning problem, because
that is actually required for a self-consistent description of the quantum vacuum.
In view of this argument, the purpose of this article is to re-consider
the zero-point energy in the
context of interacting quantum fields by exploiting non-linear equations they satisfy.

Throughout, we use natural units with $c = G = \hbar  = 1$, unless otherwise stated.

\section{Zero-point energy}

\subsection{Quantum kinetic theory}
\label{sec:qkt}

The vacuum expectation value of the
stress-energy-tensor operator $\hat{T}_\nu^\mu(x)$ of a massive non-interacting quantum
scalar field, $\hat{\phi}(x)$, in Minkowski spacetime reads
\beqa\label{eq:vev-emt}
\langle 0| \hat{T}_\nu^\mu(x) |0\rangle &=&
\frac{1}{2(2\pi)^{3}}{\int}\frac{d^3\mathbf{p}}{p_0}\,p^\mu p_\nu\,,
\eeqa
where $p_0 = (\mathbf{p}^2 + m^2)^\frac{1}{2}$ and $m$ is the scalar-field mass.
The state $|0\rangle$ denotes the Minkowski vacuum of the non-interacting
theory. This integral quartically diverges and, thereby, leads to the zero-point-energy
problem~\cite{Pauli,Zeldovich,Weinberg,Martin}.

Back in 1924, Bose suggested a phase-space description of photons
constituting black-body radiation~\cite{Bose}. It seems that we can go
one step further in this direction, by assuming that ``virtual particles" also
can be described by a distribution function. Namely, in kinetic theory,
the stress-energy tensor is defined through a distribution function
according to
\beqa\label{eq:emt-kf}
T_\nu^\mu(x) &=& {\int} \frac{d^3\mathbf{p}}{p_0}\,f(x,p)\, p^\mu p_\nu\,.
\eeqa
The function $f(x,p)$ corresponds to a distribution of $p$
at each point $x$, in the sense that $f(x,p)\,d^3\mathbf{x}\,d^3\mathbf{p}$
gives the average number of particles in the volume element $d^3\mathbf{x}$
at $\mathbf{x}$ with momenta from the interval
$(\mathbf{p},\mathbf{p} +d\mathbf{p})$~\cite{deGroot&vanLeeuwen&vanWeert}.
Leaving aside the physical interpretation of $f(x,p)$, we wish \emph{formally} to introduce a
distribution function of quantum-field fluctuations in the Minkowski vacuum, i.e.
\beqa\label{eq:vdf}
f_0(x,p) &=& \frac{1}{2(2\pi)^3}\,,
\eeqa
which, if inserted into~\eqref{eq:emt-kf}, gives~\eqref{eq:vev-emt}. This circumstance
provides a motivation for the introduction of $f_0(x,p)$.

In general, a distribution function can be defined with the help of the
covariant Wigner function
\beqa\label{eq:wf}
w_\omega(x,p) &\equiv& {\int}\frac{d^4y}{(2\pi)^4}\,e^{ipy}\,
\langle\omega| \hat{\phi}(x{+}\mbox{$\small \frac{1}{2}$}y)\hat{\phi}(x{-}\mbox{$\small \frac{1}{2}$}y)|\omega\rangle\,,
\eeqa
where $|\omega\rangle$ is a given quantum state~\cite{deGroot&vanLeeuwen&vanWeert}.\footnote{Note
that the Wigner function
is renormalised in~\cite{deGroot&vanLeeuwen&vanWeert} through the normal ordering.
We refrain from doing that, as our goal is to study the vacuum energy.}
The second moment of the Wigner function corresponds to the stress-energy tensor:
\beqa
\langle\omega| \hat{T}_\nu^\mu(x) |\omega\rangle &=& {\int}d^4p\,w_\omega(x,p)\,p^\mu p_\nu\,.
\eeqa
In particular, substituting the Wightman two-point function
\beqa
\langle 0|\hat{\phi}(x)\hat{\phi}(y)|0\rangle &=& {\int}\frac{d^3\mathbf{k}}{(2\pi)^3}\,
\frac{1}{2k^0}\,e^{-ik(x-y)}\,,
\eeqa
where $k^0 = (\mathbf{k}^2 + m^2)^\frac{1}{2}$, into~\eqref{eq:wf} with $|\omega\rangle = |0\rangle$, we find
\beqa\label{eq:vwf}
w_0(x,p) &=& \frac{1}{(2\pi)^3}\,\theta(p_0)\,\delta\big(p^2 - m^2\big)
\;\equiv\; \frac{1}{p_0}\,f_0(x,p)\,\delta\big(p_0 - (\mathbf{p}^2 + m^2)^\frac{1}{2}\big)\,,
\eeqa
where $f_0(x,p)$ agrees with~\eqref{eq:vdf}.

The Wigner function provides the phase-space description of a quantum system.
As a consequence, the distribution function can be of use to
couple various observables in local quantum field theory. Bearing in mind the standard
ultraviolet divergences in particle physics, it is tempting to conjecture that there
is a relation between them and the vacuum-energy problem.
We now wish to study how the Wigner distribution is related to
elementary particle physics and its renormalisation procedure.

\subsection{$\lambda\phi^4$-theory}
\label{sec:vdf}

The Standard Model of elementary particle physics is a theory of interacting quantum
fields~\cite{Peskin&Schroeder}. To simplify our analysis, we intend to study the simplest
non-linear field model, namely $\lambda\phi^4$-theory. This model is described by
\beqa\label{eq:a}
S[g,\phi] &=& {\int} d^4x\sqrt{-g}\,\mathcal{L}(g,\phi)\,,
\eeqa
where $g$ is a determinant of the metric tensor $g_{\mu\nu}$~\cite{Birrell&Davies},
which equals $\eta_{\mu\nu} = \text{diag}[{+}1,-1,-1,-1]$ in Minkowski spacetime, and
\beqa\label{eq:ld}
\mathcal{L}(g,\phi) &=&\frac{1}{2}\,g^{\mu\nu}\partial_\mu\phi\,\partial_\nu\phi
- \frac{1}{2}\,m_0^2 \phi^2 - \frac{\lambda_0}{4!}\,\phi^4
\eeqa
with the bare mass $m_0$ and coupling constant $\lambda_0$. Note that
a constant term can be always added to $\mathcal{L}(g,\phi)$ without changing
the scalar-field dynamics. At quantum level, this shift
of $\mathcal{L}(g,\phi)$ can be accounted for the ambiguity in working with local
products of operator-valued distributions which represent quantum
fields~\cite{Holland&Hollands}.\footnote{In this sense, the Wigner distribution is well-defined
as based on the product of quantum fields placed at different
space-time points, in contrast to its moments, due to the integration over momentum space.}

For the same reason, radiative corrections to $m_0$ and $\lambda_0$
diverge. All divergences must be regularised and then absorbed into
the non-physical parameters $m_0$, $\lambda_0$ and the field-strength renormalisation
factor $Z$ which is defined as follows:
\beqa
\hat{\phi}_r(x) &\equiv& Z^{-\frac{1}{2}}\hat{\phi}(x)\,,
\eeqa
where the index $r$ stands for ``renormalised", and
\bsubeqs\label{eq:deltas}
\beqa
\delta_Z &\equiv& Z - 1 = \sum\limits_{n = 1}^{\infty} \lambda^n\delta_Z^{(n)}\,,
\\[0.0mm]
\delta_m &\equiv& m_0^2Z - m^2 = \sum\limits_{n = 1}^{\infty} \lambda^n\delta_m^{(n)}\,,
\\[0.0mm]
\delta_\lambda &\equiv& \lambda_0 Z^2 - \lambda = \sum\limits_{n = 2}^{\infty}
\lambda^n\delta_\lambda^{(n)}\,,
\eeqa
\esubeqs
according to~\cite{Peskin&Schroeder}, where $m$ and $\lambda$ are,
respectively, the physical mass and coupling constant.

\subsubsection{Vacuum Wigner function}

In the presence of non-linear terms in the scalar-field equation, we define the
Wigner function of the Minkowski vacuum as follows:
\beqa\label{eq:vwf-lp4}
w_\Omega(x,p) &\equiv& {\int}\frac{d^4y}{(2\pi)^4}\,e^{ipy}\,
\langle\Omega| \hat{\phi}_r(x{+}\mbox{$\small \frac{1}{2}$}y)\hat{\phi}_r(x{-}
\mbox{$\small \frac{1}{2}$}y)|\Omega\rangle,
\eeqa
where $|\Omega\rangle$ denotes the Minkowski state of the interacting theory. 
Since $|\Omega\rangle$ is invariant under the Poincar\'{e} group,
the vacuum Wigner function cannot depend on the space-time point $x$ and,
therefore, we shall denote $w_\Omega(x,p)$ by $w_\Omega(p)$ in what follows.

The distribution function can be used to compute the stress tensor in the
Minkowski vacuum. Varying~\eqref{eq:a} with respect to $g_{\mu\nu}$ and
then setting $g_{\mu\nu} = \eta_{\mu\nu}$, we find
\beqa\label{eq:emt}
\hat{T}_\nu^\mu &=& \partial^\mu\hat{\phi}\,\partial_\nu\hat{\phi}
- \frac{1}{2}\,\delta_\nu^\mu\Big[(\partial\hat{\phi})^2 - m_0^2\hat{\phi}^2 -
\frac{\lambda_0}{12}\,\hat{\phi}^4\Big].
\eeqa
From this result and~\eqref{eq:vwf-lp4}, we obtain
\beqa\nonumber\label{eq:emt-lp4}
\langle\Omega| \hat{T}_\nu^\mu(x)|\Omega\rangle &=& Z{\int}d^4p\,w_\Omega(p)\,p^\mu p_\nu
\\[1mm]
&&
-\,\frac{1}{2}\delta_\nu^\mu
\left[m^2\delta_Z - \delta_m - \frac{\lambda + \delta_\lambda}{4}{\int}d^4k\,w_\Omega(k)\right]
{\int}d^4p\,w_\Omega(p)\,.
\eeqa
In the derivation of~\eqref{eq:emt-lp4}, we have taken into account that
the Minkowski vacuum is Gaussian in $\lambda\phi^4$-theory, implying that
$\langle\Omega|\hat{\phi}_r^4(x)|\Omega\rangle = 3(\langle\Omega|\hat{\phi}_r^2(x)|\Omega\rangle)^2$
generically holds, and
\beqa\label{eq:wso}
\langle\Omega|\hat{\phi}_r^2(x)|\Omega\rangle &=& {\int}d^4p\,w_\Omega(p)\,,
\eeqa
which directly follows from~\eqref{eq:vwf-lp4}.

Next, we find from the scalar-field equation,
\beqa
\left[Z\big(\Box_x + m_0^2\big) + \frac{\lambda_0Z^2}{3!}\,\hat{\phi}_r^2(x)\right]\hat{\phi}_r(x) &=& 0\,,
\eeqa
that
\beqa\label{eq:bem}
\left[Z\big(p^2 - m_0^2\big) - \frac{\lambda_0Z^2}{2}{\int}d^4k\,w_\Omega(k)\right]w_\Omega(p)
&=& 0\,,
\eeqa
where we have taken into consideration that the state $|\Omega\rangle$ is Lorentz invariant
and Gaussian, providing for
$\langle\Omega|\hat{\phi}_r(y)\hat{\phi}_r^3(x)|\Omega\rangle =
3\langle\Omega|\hat{\phi}_r(y)\hat{\phi}_r(x)|\Omega\rangle\langle\Omega|\hat{\phi}_r^2(x)|\Omega\rangle$
(cf.~(14) in Sec.~III.2 of~\cite{deGroot&vanLeeuwen&vanWeert}).
Integrating~\eqref{eq:bem} over $p$ and then substituting into~\eqref{eq:emt-lp4}, we
find
\beqa
\langle\Omega| \hat{T}_\mu^\mu(x) |\Omega\rangle &=& {\int}d^4p\,w_\Omega(p)
\big[2(p^2 - m^2)Z + m_0^2Z\big].
\eeqa
Alternatively, taking the trace of~\eqref{eq:emt} and then using the scalar-field
equation to eliminate the quartic term, we obtain
\beqa
\langle\Omega| \hat{T}_\mu^\mu(x) |\Omega\rangle &=&
m_0^2Z \langle\Omega|\hat{\phi}_r^2(x)|\Omega\rangle\,,
\eeqa
where we have employed
$\langle\Omega|\partial\hat{\phi}_r\partial\hat{\phi}_r + \hat{\phi}_r\Box\hat{\phi}_r|\Omega\rangle = 0$,
coming from
$\partial\hat{\phi}_r = i[\hat{P},\hat{\phi}_r]$,
where $\hat{P}$ is the space-time translation operator, and $\hat{P}|\Omega\rangle = P|\Omega\rangle$,
where $P \in \mathbb{R}^4$.
Comparing these traces with~\eqref{eq:wso}
taken into account, we find that
$\big(p^2 - m^2\big)w_\Omega(p) = 0$. This results in
\beqa\label{eq:wrp}
w_\Omega(p) &=& \frac{1}{(2\pi)^3}\,\theta(p_0)\,\delta\big(p^2 - m^2\big)\,\delta_w\,.
\eeqa
The as-yet-unknown parameter\marginnote{R6}
\beqa
\delta_w &\equiv& \sum\limits_{n = 0}^{\infty}\lambda^n \delta_w^{(n)}
\eeqa
needs to be determined, where, however, $\delta_w^{(0)} = 1$ must hold as
it follows from~\eqref{eq:vwf}.

Now, substituting $w_\Omega(p)$
in~\eqref{eq:bem} and then integrating over $p$, we find the main equation
of this section:
\beqa\label{eq:main-eq}
\big[m^2\delta_Z - \delta_m\big]{\int}d^4p\,w_\Omega(p) &=& \frac{\lambda +\delta_\lambda}{2}
\left[{\int}d^4p\,w_\Omega(p)\right]^2\hspace{-1mm}.
\eeqa
Note,~\eqref{eq:main-eq} does not exist if $\lambda = 0$. But, since $\lambda \neq 0$,
we have two solutions of this equation, namely
\bsubeqs\label{eq:sol1-sol2}
\beqa\label{eq:sol2}
\frac{1}{2}\,{\int}d^4p\,w_\Omega(p)  &=& 0\,,
\\[0.5mm]\label{eq:sol1}
\frac{1}{2}\,{\int}d^4p\,w_\Omega(p) &=&
\frac{m^2\delta_Z - \delta_m}{\lambda + \delta_\lambda}\,.
\eeqa
\esubeqs
The trivial solution~\eqref{eq:sol2} corresponds to the absence of the (regularised) zero-point energy,
while the non-trivial solution~\eqref{eq:sol1} implies that the vacuum energy is non-zero.

We find from~\eqref{eq:wrp} and~\eqref{eq:sol1} that
\bsubeqs\label{eq:FR0-FR1}
\beqa\label{eq:Fr0}
\delta_w^{(0)} &=&
\frac{2(4\pi)^\frac{d}{2}}{(m^2)^{\frac{d}{2}-1}}\frac{m^2\delta_Z^{(1)} {-} \delta_m^{(1)}}
{\Gamma\big(1{-}\mbox{\small $\frac{d}{2}$}\big)}\,,
\\[2mm]\label{eq:Fr1}
\delta_w^{(1)} &=& \frac{2(4\pi)^\frac{d}{2}}{(m^2)^{\frac{d}{2}-1}}
\frac{m^2\delta_Z^{(2)} {-} \delta_m^{(2)} {-} \big[m^2\delta_Z^{(1)} {-}\delta_m^{(1)}\big]
\delta_\lambda^{(2)}}
{\Gamma\big(1{-}\mbox{\small $\frac{d}{2}$}\big)}
\eeqa
\esubeqs
and so forth, where we have used dimensional regularisation, $4 \rightarrow d < 4$, on the left-hand
side of~\eqref{eq:sol1}.
It follows from~\eqref{eq:FR0-FR1} that $\delta_w$ can be determined
by computing $\delta_Z$, $\delta_m$ and $\delta_\lambda$. These come in turn from the
renormalisation of self-energy and vertex diagrams~\cite{Peskin&Schroeder}.

\subsubsection{Self-energy renormalisation}
\label{sec:ser}

The Feynman propagator gets loop corrections which change its pole structure.
We can represent this circumstance pictorially as follows:
\beqa
\mathbf{\imineq{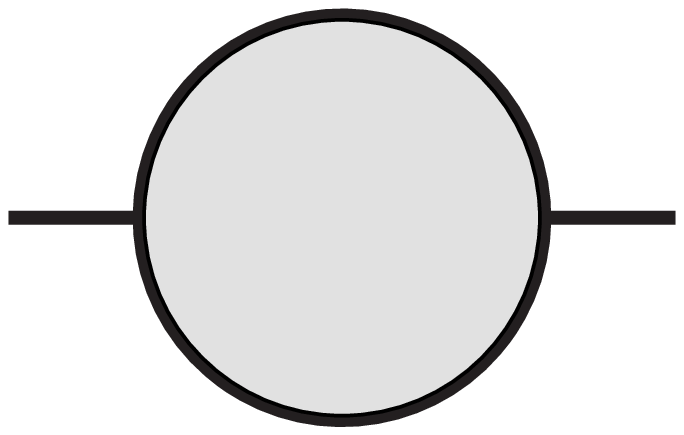}{5.5}} &=& \frac{i}{p^2 - m^2 - M^2(p)}\,,
\eeqa
where the shaded circle denotes the sum of all possible (one-particle irreducible) self-energy diagrams.
According to the standard renormalisation conditions~\cite{Peskin&Schroeder},
$M^2(p)$ satisfies
\bsubeqs\label{eq:serc}
\beqa\label{eq:serc-1}
M^2(p)\big|_{p^2 = m^2} &=& 0\,,
\\[1mm]\label{eq:serc-2}
\frac{d}{dp^2}\,M^2(p)\big|_{p^2 = m^2} &=& 0\,.
\eeqa
\esubeqs
These conditions turn out to be sufficient to compute $\delta_w$ up to the linear
order in the coupling constant $\lambda$.

Specifically, at one-loop order, we have
\beqa\label{eq:sed1l}
-iM^2(p) &=& \mathbf{\imineq{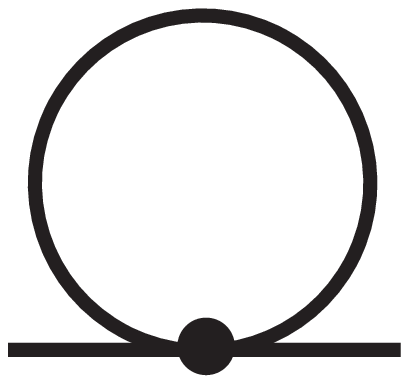}{5.5}} + \mathbf{\imineq{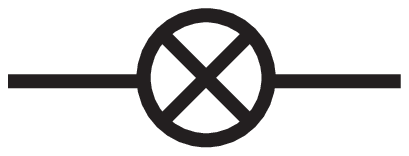}{2.2}}\,,
\eeqa
where the counter-term vertex
\beqa
\mathbf{\imineq{counterterm-1.eps}{2.2}} &=& i\big(p^2\delta_Z - \delta_m\big)\,.
\eeqa
It is straightforward to compute the first diagram in~\eqref{eq:sed1l}, which gives
from the renormalisation conditions~\eqref{eq:serc}
the following results (cf.~(10.30) in~\cite{Peskin&Schroeder}):
\bsubeqs\label{eq:dz1-dm1}
\beqa
\delta_m^{(1)} &=& - \frac{1}{2}\frac{(m^2)^{\frac{d}{2}-1}}{(4\pi)^\frac{d}{2}}\,
\Gamma\big(1{-}\mbox{\small $\frac{d}{2}$}\big)\,,
\\[1mm]
\delta_Z^{(1)} &=& 0\,.
\eeqa
\esubeqs
Substituting~\eqref{eq:dz1-dm1} into~\eqref{eq:Fr0}, we obtain
\beqa\label{eq:Fr0v}
\delta_w^{(0)} &=& 1\,.
\eeqa

At two-loop order, however, we find
\beqa\label{eq:sed2l}
-iM^2(p) &=& \mathbf{\imineq{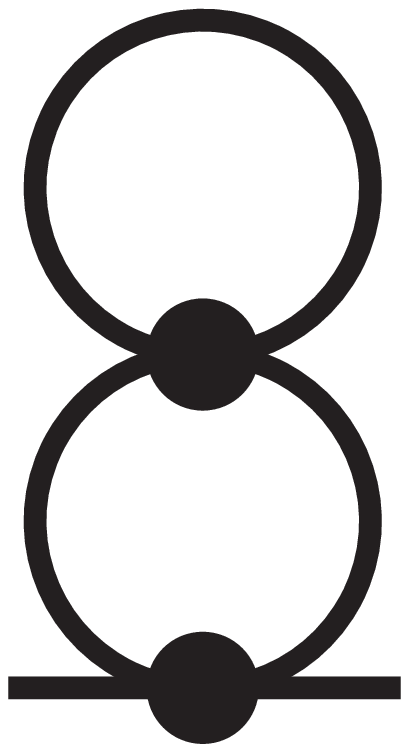}{5.5}}\, +
\mathbf{\imineq{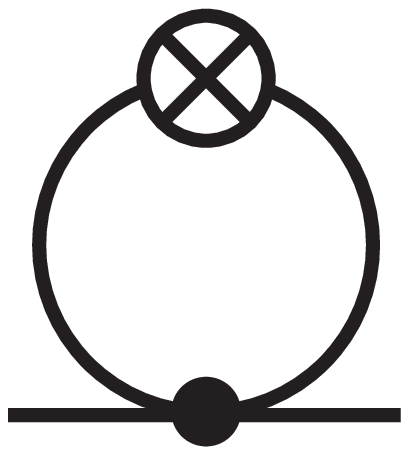}{5.5}} +
\mathbf{\imineq{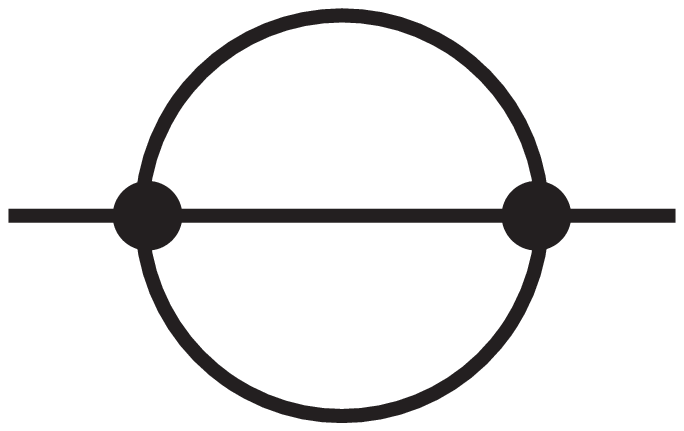}{5.5}} + \mathbf{\imineq{counterterm-1.eps}{2.2}}
+ \mathbf{\imineq{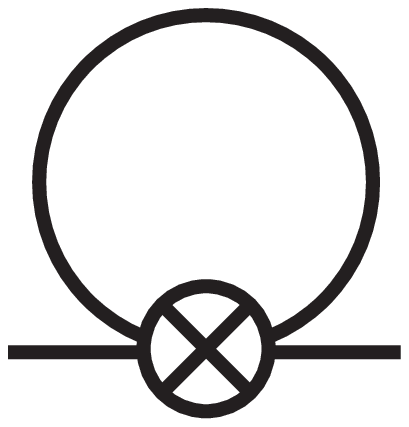}{5.5}}\,,
\eeqa
where the counter-term vertex
\beqa
\mathbf{\imineq{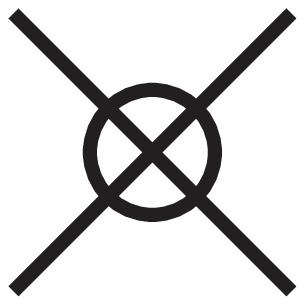}{4.0}} &=& -i\delta_\lambda\,.
\eeqa
The first two diagrams in~\eqref{eq:sed2l} cancel each other, whereas
the last three diagrams give
\beqa\label{eq:two-loop-diag}
\left[\mathbf{\imineq{2-loop-2.eps}{5.5}}\right] &=&
-i\lambda^2\Big(m^2\delta_Z^{(2)}-\delta_m^{(2)} + \delta_m^{(1)}\delta_\lambda^{(2)}\Big),
\eeqa
where the renormalisation condition~\eqref{eq:serc-1} has been taken into account and
the square brackets mean that the diagram is taken on the mass shell, i.e. $p^2 = m^2$.
Making use of~\eqref{eq:dz1-dm1} and~\eqref{eq:two-loop-diag}, we obtain from~\eqref{eq:Fr1}
that
\beqa\label{eq:Fr1v}
\delta_w^{(1)} &=& \frac{2(4\pi)^\frac{d}{2}}{(m^2)^{\frac{d}{2}-1}}
\frac{i/\lambda^2}
{\Gamma\big(1{-}\mbox{\small $\frac{d}{2}$}\big)}\left[\mathbf{\imineq{2-loop-2.eps}{5.5}}\right] \;\neq\; 0\,.
\eeqa

\subsubsection{Vacuum bubbles}
\label{sec:zpe}

From another side, the parameter $\delta_w$ can be determined by considering
vacuum-bubble diagrams. In general, we have
\beqa\label{eq:vacuum-energy}
\hspace{-6mm}
-i\langle\Omega| \hat{T}_0^0(x) |\Omega\rangle
&=& \mathbf{\imineq{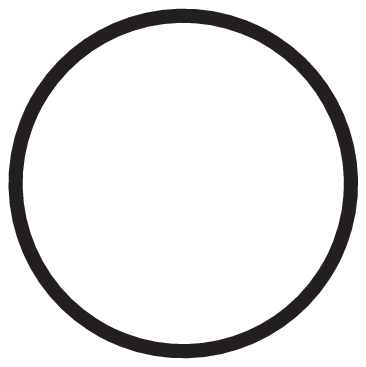}{5.5}} + \mathbf{\imineq{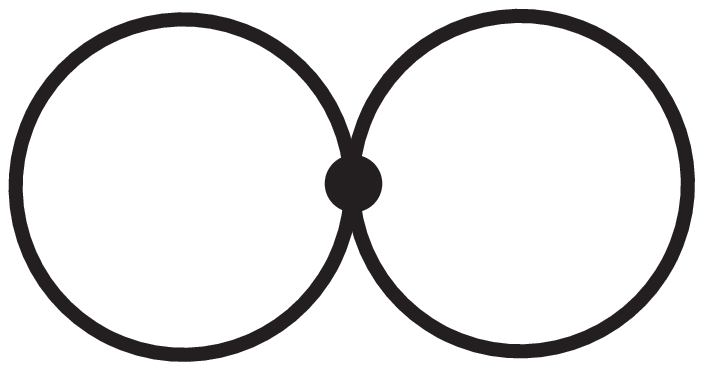}{5.5}}
+ \mathbf{\imineq{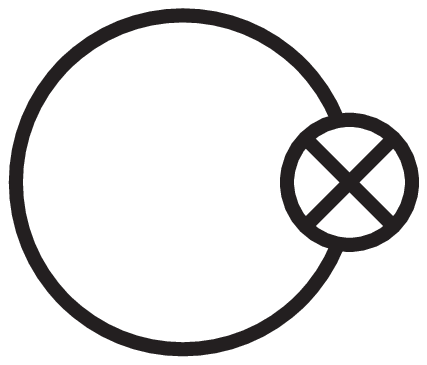}{5.5}} + \text{O}\big(\lambda^2\big)\,.
\eeqa
Up to the linear order in the coupling constant $\lambda$, we find
from~\eqref{eq:emt-lp4} with~\eqref{eq:wrp} that
\beqa\nonumber\label{eq:one-loop-vbd}
\langle\Omega| \hat{T}_0^0(x) |\Omega\rangle &=& \frac{1}{d}\frac{(m^2)^\frac{d}{2}}{(4\pi)^\frac{d}{2}}
\Gamma\big(1{-}\mbox{\small $\frac{d}{2}$}\big)
\big(\delta_w^{(0)} + \lambda\delta_w^{(1)}\big)
\\[1mm]
&&
-\,\frac{\lambda}{8}\frac{(m^2)^{d-2}}{(4\pi)^d}\Gamma^2\big(1{-}\mbox{\small $\frac{d}{2}$}\big)
\big(2-\delta_w^{(0)}\big)\,\delta_w^{(0)}
+ \text{O}\big(\lambda^2\big),
\eeqa
where~\eqref{eq:dz1-dm1} have been used.

Computing the first diagram in~\eqref{eq:vacuum-energy}, we find
\beqa\label{eq:fr0v}
\delta_w^{(0)} &=& 1\,,
\eeqa
which is consistent with the result~\eqref{eq:Fr0v}. The second and third diagrams shown
in~\eqref{eq:vacuum-energy} give
\beqa\label{eq:fr1v}
\delta_w^{(1)} &=& 0\,,
\eeqa
which does not agree with the result~\eqref{eq:Fr1v}.

\subsection{Discussion}

The tension between~\eqref{eq:Fr1v} and~\eqref{eq:fr1v} is not an artifact
of dimensional regularisation. In fact, it is straightforward to show that the
same problem also takes place at two-loop order in Pauli-Villars regularisation.

To summarise, it follows from the non-vanishing value of the vacuum energy density that
the Wigner function of the Minkowski state, $w_\Omega(p)$, is non-zero. But,
the substitution of $w_\Omega(p)$ into~\eqref{eq:main-eq} does not solve this equation
at two-loop order in perturbation theory. Thus, $w_\Omega(p) \neq 0$ violates
the scalar-field equation. Since this is the main equation describing the quantum-field
dynamics, its violation is unacceptable. According to~\eqref{eq:main-eq}, the only option left
is to choose its trivial solution~\eqref{eq:sol2}.
Consequently, $w_\Omega(x,p)$ defined in~\eqref{eq:vwf-lp4} has to be
renormalised to zero.
This implies in turn that the Minkowski vacuum has neither energy nor pressure.

Alternatively, we may consider
$\mathcal{L}(g,\phi) \rightarrow \mathcal{L}(g,\phi) - \Lambda_0/8\pi$,
where $\mathcal{L}(g,\phi)$ is defined in~\eqref{eq:ld} and $\Lambda_0$ is a
constant which cancels the sum of all vacuum-bubble diagrams in the
scalar-field model. This procedure does not pose a fine-tuning problem, as that
is a self-consistency condition.

As emphasised above, the crucial equation~\eqref{eq:main-eq}, which results
in the tension between the renormalisation conditions~\eqref{eq:serc}
and the non-renormalised Wigner function, arises from the scalar-field
equation. We have studied $\lambda\phi^4$-theory so far, but the
same conclusion holds for other non-linear models. For example, the field
equation in $g\phi^3$-theory gives
\beqa\label{eq:main-eq-phi3}
\frac{g_0Z^\frac{3}{2}}{2}{\int}d^4p\,w_\Omega(p) &=& 0\,.
\eeqa
We are again forced to renormalise the Wigner
function here, in such a way~\eqref{eq:main-eq-phi3} is trivially satisfied.

The microscopic effects due to quantum-field fluctuations are
observable only in the presence of matter. The fact that the renormalised
vacuum Wigner function has to vanish does not lead to any tension with
the up-to-date observations. It is because vacuum-bubble diagrams
do not have external legs and, for this reason, these diagrams are not coupled
through non-gravitational fields to elementary particles or atoms.

\section{Conclusion}

The Universe we observe is not Minkowski spacetime, although we have made this
approximation as it is common in elementary particle physics~\cite{Peskin&Schroeder}.
In particular, the expectation value of~\eqref{eq:emt} in a given quantum
state represents a source term in the Einstein field equations, which distorts the
spacetime. Still, the Universe locally looks as Minkowski spacetime, in accordance with
the Equivalence Principle. Our computations are therefore valid in any local
Lorentz frame. The conclusion is then that the Minkowski quantum
vacuum in elementary particle physics is not a source of the gravitational field.

In curved spacetime, the vacuum expectation value of
the stress-tensor operator acquires, in general, curvature-dependent corrections.
On dimensional grounds, the first non-trivial covariantly-conserved correction has the form
\beqa\label{eq:emt-cdc}
G_\nu^\mu(x)\langle \Omega|\hat{\phi}^2(x)|\Omega\rangle\,,
\eeqa
where $G_\nu^\mu(x)$ is the Einstein tensor and we have taken into account that
$\langle \Omega|\hat{\phi}^2(x)|\Omega\rangle$ does not depend on $x$, due to
local homogeneity of the Minkowski state. 
This correction quadratically diverges, while higher-order curvature corrections
are at worst logarithmically divergent. Hence,~\eqref{eq:emt-cdc} leads to an infinite shift of the inverse
gravitational constant, $1/G$. This shift coincides up to a numerical factor
with the result from effective-action computations (see, e.g., Sec.~6.2 in~\cite{Birrell&Davies}).
However, according to the self-consistency argument, the
renormalised Wigner function is zero, which, according to~\eqref{eq:wso},
implies that~\eqref{eq:emt-cdc} vanishes.

Yet, the very fact that the Minkowski vacuum does not gravitate
in elementary particle physics does not solve the main cosmological constant problem.
Indeed, the Higgs condensate contributes a negative energy density of order
$(100\,\text{GeV})^4$ to the cosmological constant~\cite{Weinberg}. The energy density of dark energy
is, however, of order $(0.001\,\text{eV})^4$. Besides, it seems that
the vacuum in quantum chromodynamics makes a contribution of order $(0.3\,\text{GeV})^4$ to
the total vacuum energy of the Universe~\cite{Martin}. The former contribution may be got rid
of by re-defining the Higgs potential, while the latter, probably, requires novel ideas,
such as~\cite{Holland&Hollands,Klinkhamer&Volovik}, to harmonise
the Standard Model with astrophysical observations.

Admittedly, the physical meaning of the distribution function of ``virtual particles"
is not entirely clear. In this article, we have used this concept as a mathematical tool to
study physics of the zero-point energy. We hope to come back to this question
later.

\section*{
ACKNOWLEDGMENTS}

It is a pleasure to thank Frans Klinkhamer for valuable discussions. I am also
grateful to Claus Kiefer, Roman Lee and Stefan Liebler for discussions.


\begin{thebibliography}{99}

\bibitem{Pauli}
W.\,E. Pauli, in
\emph{Exclusion principle and quantum mechanics}
(Nobel lecture, 1946).

\bibitem{Zeldovich}
Ya.\,B. Zeldovich, Sov. Phys. Usp. {\bf 11} (1968) 381.

\bibitem{Weinberg}
S. Weinberg, Rev. Mod. Phys. {\bf 61} (1989) 1.

\bibitem{Martin}
J. Martin, C.\,R. Phys. {\bf 13} (2012) 566.

\bibitem{Milonni}
P.\,W. Milonni,
\emph{The Quantum Vacuum. An Introduction to Quantum Electrodynamics}
(Academic Press, Inc., 1994).

\bibitem{Bose}
S.\,N. Bose, Zeitschrift f\"{u}r Physik {\bf 26} (1924) 178.

\bibitem{deGroot&vanLeeuwen&vanWeert}
S.\,R. de Groot, W.\,A. van Leeuwen, Ch.\,G. van Weert,
\emph{Relativistic Kinetic Theory. Principles and Applications}
(North-Holland Publishing Co., 1980).

\bibitem{Peskin&Schroeder}
M.\,E.\, Peskin, D.\,V.\, Schroeder,
\emph{An Introduction to Quantum Field Theory}
(Addison-Wesley Publishing Co., 1995).

\bibitem{Birrell&Davies}
N.\,D.\, Birrell, P.\,C.\,W.\, Davies,
\emph{Quantum Fields in Curved Space}
(Cambridge University Press, 1984).

\bibitem{Holland&Hollands}
J. Holland, S. Hollands, Class. Quantum Grav. {\bf 31} (2014) 125006.

\bibitem{Klinkhamer&Volovik}
F.\,R. Klinkhamer, G.\,E. Volovik, Phys. Rev. D {\bf 77} (2008) 085015.

\end{thebibliography}
\end{document}